\documentclass[11pt]{article}
\usepackage[dvips]{graphicx}
\usepackage{psfrag}

\makeatletter
\def\txtline#1{\noalign{\hbox{\strut\hskip\@totalleftmargin {#1}}}}
\newcommand{\bra}[1]{\langle #1|}
\newcommand{\ket}[1]{|#1\rangle}
\makeatother
\newcommand{\zero}{\mbox{$\ket 0$}}
\newcommand{\one}{\mbox{$\ket 1$}}
\newcommand{\zz}{\mbox{$\ket{00}$}}
\newcommand{\zo}{\mbox{$\ket{01}$}}
\newcommand{\oz}{\mbox{$\ket{10}$}}
\newcommand{\oo}{\mbox{$\ket{11}$}}

\newcommand{\squash}[1]{\raisebox{0.1ex}[0pt][0pt]{\small$\textstyle #1$}}
\newcommand{\oosrt}{\squash{\frac{1}{\sqrt{2}}}}
\newcommand{\iosrt}{\squash{\frac{i}{\sqrt{2}}}}
\newcommand{\mystrut}{\raisebox{2mm}{\rule[-5mm]{0mm}{9mm}}}
\newcommand{\mxstrut}{\raisebox{2mm}{\rule[-5mm]{0mm}{6mm}}}

\newcommand{\pbfrac}[2]{\mbox{$\mbox{}^{#1}\!/_{#2}$}}
\newcommand{\ident}{\textup{I}}
\newcommand{\WH}{\textsf{H}}
\newcommand{\WHs}{\textsf{\scriptsize H}}
\renewcommand{\epsilon}{\varepsilon}
\renewcommand{\phi}{\varphi}
\renewcommand{\log}{\lg}

\newtheorem{theorem}{Theorem}
\newtheorem{proposition}[theorem]{Proposition}

\newtheorem{corollary}[theorem]{Corollary}

\newenvironment{proof}{
\par
\noindent {\textbf{Proof.}}\rm}%
{\hspace*{\fill}\rule{0.5em}{0.809em}\par}

\begin{document}

\title{Quantum Computing Without Entanglement\,%
\thanks{The work of E.B., D.K. and T.M.~is supported 
in parts by the Israel MOD Research and Technology Unit. 
The work of E.B.~is supported in
parts also by the fund for the promotion of research at the Technion, 
and by the European Commission through the IST Programme under contract
IST-1999-11234.
The work of G.B. is supported in parts by
Canada's NSERC, Qu\'ebec's FCAR, The Canada Research Chair Programme
and the Canadian Institute for Advanced Research.
}}

\author{Eli Biham$^{\,1}$, Gilles Brassard$^{\,2}$, Dan Kenigsberg$^{\,1}$
and Tal Mor$^{\,1\,\dagger}$
\\ \small 1. Computer Science Department, Technion, 
Haifa 32000, Israel.
\\ \small 2. DIRO, Universit\'e de Montr\'eal, Montr\'eal, Qu\'ebec,
H3C 3J7 Canada.
\\ \small $\dagger$ To whom correspondence should be addressed. 
Email: talmo@cs.technion.ac.il}

\date{25 June 2003}    

\maketitle
\begin{abstract}

It is generally believed that entanglement is essential for quantum computing.
We~present here a few simple examples 
in which quantum computing \emph{without}
entanglement is better than anything classically achievable,
in terms of the reliability of the outcome after a fixed number of oracle calls.
Using a separable (that is, unentangled) 
$n$-qubit state,
we show that 
the Deutsch-Jozsa problem and the Simon problem can be solved
more reliably by a quantum computer 
than by the best possible classical algorithm, even probabilistic.
We conclude that:
(a)~entanglement is not essential for quantum computing; and
(b)~some advantage of quantum algorithms 
over classical algorithms persists even when
the quantum state contains an arbitrarily small amount of information---that
is,  even when the state is arbitrarily close to being totally mixed.
\end{abstract}

\section{Introduction}

Quantum computing is a new fascinating field of research in which 
the rules of quantum mechanics are used to solve various computing
problems more efficiently than any classical algorithm could
do~\cite{Nielsen00, Gruska98}.
This has been rigourously demonstrated in oracle settings~\cite{oracle}
and there is
significant evidence that it is true in unrelativized cases as well~\cite{Shor}.
The quantum unit of information is called the \emph{qubit}.
In addition to the ``classical'' states
$|0\rangle$ and $|1\rangle$, a qubit can be in any superposition
$\ket\psi = \alpha |0\rangle + \beta |1\rangle$, where $| \cdot \rangle$ is the
standard Dirac notation for quantum states, and $\alpha$ and $\beta$ are complex
numbers subject to $|\alpha|^2 + |\beta|^2 = 1$.
For instance, $\ket\pm=\oosrt\zero\pm\oosrt\one$ 
and $\ket{\pm_i}=\oosrt\zero\pm \iosrt\one$ are some specific
pure states that we shall use later on.
When $n$ qubits are used, their state can be 
in a superposition of all ``classical'' $n$-bit states,
that is, $|\psi\rangle=\sum_{i=0}^{2^n - 1} \alpha_i |i\rangle$, where $i$ 
is written in binary representation and $\sum_i |\alpha_i|^2 = 1 $.
These states are called \emph{pure states}. 

If qubits $\ket{\psi}$ and $\ket{\phi}$ are in states
\mbox{$\alpha\zero+\beta\one$} and \mbox{$\gamma\zero+\delta\one$}, respectively, 
the state of a two-qubit system composed of those two qubits is given by their
\emph{tensor product}
\begin{equation}\label{eq:tensor}
\ket{\psi}\otimes\ket{\phi} = \alpha\gamma\zz + \alpha\delta\zo 
   + \beta\gamma\oz + \beta\delta\oo \, .
\end{equation}
This notion generalizes in the obvious way to the tensor product of arbitrarily many
quantum systems.
Perhaps the most nonclassical aspect of quantum information processing
stems from the fact that not all two-qubit states can be written in the
form of Eq.~(\ref{eq:tensor}).
For instance, the states
$\ket{\Psi^{\pm}} = \oosrt\ket{01} \pm \oosrt\ket{10}$
and 
$\ket{\Phi^{\pm}} = \oosrt\ket{00} \pm \oosrt\ket{11}$,
known as the \emph{Bell states}---or~perhaps
more appropriately the Braunstein-Mann-Revzen (BMR)
states~\cite{BMR}---do not factor out as a tensor product.
In general, state
\mbox{$a\zz+b\zo+c\oz+d\oo$} can be written in the form
of Eq.~(\ref{eq:tensor}) if and only if \mbox{$ad=bc$}.
Multiple-qubits pure states that can 
be written as a tensor product of the individual qubits
are said to be \emph{separable}, or \emph{product} states.
Otherwise they are \emph{entangled}.

When information is lacking about the state of a qubit,
we say that this qubit is in a \emph{mixed state}. 
This is described by a matrix $\rho = \sum_j p_j |\psi_j \rangle \langle
\psi_j |$, called the \emph{density matrix},
with $p_j$ being the probability of each (pure) state
$|\psi_j\rangle$. 
The representation of $\rho$ as a sum is not unique.
For instance, an equal mixture of 
$\ket{\psi_+(\theta)}=\cos\theta\ket0+\sin\theta\ket1$ and
$\ket{\psi_-(\theta)}=\cos\theta\ket0-\sin\theta\ket1$ 
is written as
$\rho_{\theta}=\frac12 \ket{\psi_+(\theta)}\bra{\psi_+(\theta)}
+ \frac12 \ket{\psi_-(\theta)}\bra{\psi_-(\theta)}$.
Simple algebra shows that this is in fact exactly the \emph{same} as
$(\cos^2\theta)\ket0\bra0+
(\sin^2\theta)\ket1\bra1$,
which is in general an unequal mixture of $\zero$ and~$\one$.
A quantum mixed state $\rho$ of several qubits
is called a \emph{product state}
if it can be written as a tensor product of
the states of the individual qubits,
such as \mbox{$\rho = \rho_A \otimes \rho_B$}.

Recall that in the case of pure states, any product state is separable
and any non-product state is entangled.  
The situation with mixed states
is different---and more interesting---because
there are separable states that are not product states.
We say that a multiple-qubit
mixed state is \emph{separable} if it can be written as
$\rho = \sum_j p_j |\psi_j \rangle \langle
\psi_j |$ such that each of the $|\psi_j \rangle$ is a
separable pure state.
Equivalently, a mixed state $\rho$ is separable if it
can be written in the form $\rho = \sum_j p_j \rho_j$ 
such that each of the $\rho_j$ is a product state.

The intuition behind this definition is that a state (pure or mixed)
is separable if and only if it can be prepared in remote locations with
the help of classical communication only.
For instance, Alice and Bob can remotely prepare the separable 
bipartite state
\mbox{\(\frac12\ket{01}\bra{01}+\frac12\ket{10}\bra{10}\)} as follows.
Alice tosses a fair coin and tells the outcome 
to Bob over a classical channel.
If~the coin came up heads, Alice and Bob prepare their qubits in states
\zero\ and \one, respectively.
But if the coin comes up tails, they prepare their qubits in
states \one\ and \zero.
Then, \emph{provided Alice and Bob forget the outcome of the coin},
they are left with the desired state.

If~a mixed state is not separable, 
then we say that it is \emph{entangled}.
Deciding if a mixed state is entangled or separable is not
an easy task in the general case because its representation is
not unique.
For instance, the state  
\(\frac12\ket{\Psi^+}\bra{\Psi^+}+\frac12\ket{\Psi^-}\bra{\Psi^-} \)
is separable, despite being a mixture of two entangled pure states,
because it can be written equivalently as
\(\frac12\ket{01}\bra{01}+\frac12\ket{10}\bra{10}\)\@.
As a more sophisticated example, a \emph{Werner state}~\cite{Werner_ref}
$\chi = \lambda |\Psi^-\rangle \langle\Psi^-|
+ \frac{1-\lambda}{3}\left[ \,
|\Psi^+\rangle \langle\Psi^+| +
|\Phi^-\rangle \langle\Phi^-| +
|\Phi^+\rangle \langle\Phi^+| \, \right] $, which can also be written as 
\begin{equation}
\chi = \epsilon|\Psi^-\rangle \langle\Psi^-|
+ (1-\epsilon) \, \ident / 4 \label{Werner}
\end{equation}
with $\epsilon= (4\lambda - 1)/3$ and $\ident$ a $4 \times 4$ identity
matrix, is entangled if and only if $\lambda > \frac12$, or equivalently 
$\epsilon > \frac13$.
For~$\lambda=\frac14$ ($\epsilon=0$) the state is fully mixed, hence it
contains no information.
For $\lambda=\frac12$ ($\epsilon=\frac13$) the state can be
rewritten as
\[\textstyle \frac16(\ket{\Psi^+}\bra{\Psi^+}+\ket{\Psi^-}\bra{\Psi^-})+
\frac16(\ket{\Phi^-}\bra{\Phi^-}+\ket{\Psi^-}\bra{\Psi^-})+
\frac16(\ket{\Phi^+}\bra{\Phi^+}+\ket{\Psi^-}\bra{\Psi^-}) \, ,\]
which makes its separability immediately apparent because
\[\mbox{$\ket{\Phi^-}\bra{\Phi^-}+\ket{\Psi^-}\bra{\Psi^-}$}~=~
\mbox{$\ket{+-}\bra{+-}+\ket{-+}\bra{-+}$}\] and
\[\mbox{$\ket{\Phi^+}\bra{\Phi^+}+\ket{\Psi^-}\bra{\Psi^-}$}~=~
\mbox{$\ket{+_i-_i}\bra{+_i-_i}+\ket{-_i+_i}\bra{-_i+_i}$} \, . \]
Note that, although separable, this state is far from being classical,
as only a nontrivial mixture of states written in different bases
exposes its separability.

\emph{Quantum computers} can manipulate quantum information
by means of unitary transformations~\cite{Gruska98,Nielsen00,Deutsch89}.
In~particular, they can work with superpositions.
For instance,
a single-qubit Walsh--Hadamard operation $\WH$ transforms a qubit from $\ket0$
to $\ket+$ and from $\ket1$ to $\ket-$. When $\WH$ is applied to
a superposition such as $\ket+$,
it follows by the linearity of quantum mechanics that the resulting
state is $\frac12((\ket0+\ket1)+(\ket0-\ket1))=\ket0$.
This illustrates the phenomenon of destructive \emph{interference},
by which component $\ket1$ of the state is erased.
Consider now an $n$-qubit quantum register initialized to~$|0^n\rangle$.
Applying a Walsh--Hadamard transform to each of these qubits
yields an equal superposition of all $n$-bit classical states:
\[ |0^n\rangle \stackrel \WHs\longrightarrow
\frac1{2^{n/2}}\sum_{x=0}^{2^n-1} |x\rangle \, . \]

Consider now a function $f:\{0,1\}^n \longrightarrow \{0,1\}$ that
maps $n$-bit strings to a single bit.  On a quantum computer,
because unitary transformations are \emph{reversible}, it is
natural to implement it as a unitary transformation $U_f$ that maps
$\ket{x}\ket{b}$ to $\ket{x}\ket{b\oplus f(x)}$, where $x$ is an $n$-bit
string, $b$ is a single bit, and ``$\oplus$'' denotes the exclusive-or.
Schematically,

\begin{equation} 
|x\rangle |b\rangle \stackrel{U_f}\longrightarrow |x\rangle | f(x) \oplus b \rangle
\, .
\label{eq:oracle}
\end{equation}
The linearity of quantum mechanics gives rise to two important phenomena.
(1)~\emph{Quantum parallelism}: we can compute $f$ on arbitrarily many classical
inputs by a single application of $U_f$ to a suitable superposition:
\begin{equation} 
\sum_x \alpha_x |x\rangle |b\rangle \stackrel{U_f}\longrightarrow
\sum_x \alpha_x |x\rangle | f(x) \oplus b \rangle
\, .
\end{equation}
When this is done, the additional output qubit may become entangled
with the input register;
(2)~\emph{Phase kick-back}: the outcome of $f$ can be recorded in the \emph{phase}
of the input register rather than being XOR-ed to the additional output qubit:
\begin{equation} 
|x\rangle |-\rangle \stackrel{U_f}\longrightarrow (-1)^{f(x)}|x\rangle | - \rangle
\, ;~~~\
\sum_x \alpha_x \ket x \ket- \stackrel{U_f}\longrightarrow
\sum_x \alpha_x (-1)^{f(x)}\ket x \ket-
\, .\end{equation}

Much of the current interest in quantum computation was spurred by Peter Shor's
momentous discovery that quantum computers can in principle factor large numbers
and extract discrete logarithms
in polynomial time~\cite{Shor} and thus break much of contemporary
cryptography, such as the RSA cryptosystem and the Diffie-Hellman public-key
exchange. 
However, this does not provide a proven advantage of quantum
computation because nobody knows for sure that these problems are 
genuinely
hard for classical computers.  
On the other hand, it has been demonstrated that
quantum computers can solve some problems exponentially faster than any
classical computer provided the input is given as an
\emph{oracle}~\cite{DJ92,oracle}, 
and even if we allow bounded errors~\cite{Simon97}.
In~this model, some function $f:\{0,1\}^n \rightarrow \{0,1\}$ is given
as a black-box, which means that the only way to obtain knowledge about $f$
is to query the black-box on chosen inputs.  In the corresponding quantum
oracle model, a function $f$ is provided by a black-box that applies
unitary transformation $U_f$ to any chosen quantum state, as described by
Eq.~(\ref{eq:oracle}).
The goal of the algorithm is to learn some property
of the function.

A fundamental question is: Where does the surprising computational advantage 
provided by quantum mechanics come from? What is the
nonclassical property of quantum mechanics that leads to such an
advantage? Do superposition and interference provide
the quantum advantage?
Probably the most often heard answer is that 
the power of quantum computing
comes from the \textbf{use of entanglement},
and indeed there are very strong arguments in favour of this belief.
(See~\cite{JozsaLinden,braunstein-et-al,LindenPopescu01,
EkertJozsa98,SchackCaves99} for a discussion.)

We show in this paper that this common belief is \textbf{wrong}.
To this effect, we present two simple examples
in which quantum algorithms are better than classical algorithms
even when no entanglement is present.
Furthermore, we show that quantum algorithms can be better 
than classical algorithms even when the state of the computer is
almost totally mixed---which means that it contains an arbitrarily small
amount of information.

The most usual measure of efficiency for computer algorithms is the amount
of time required to obtain the solution, as function of the input size.
In oracle setting this usually means the number of queries needed to gain a
predefined amount of information about the solution.
Here, we fix a maximum number of oracle calls and we try to obtain
as much Shannon information as possible about the correct answer.
In this model, we analyse two famous problems due to
Deutsch-Jozsa~\cite{DJ92} and Simon~\cite{Simon97}.
We show that, when a single oracle query is performed, 
the probability to
obtain the correct answer is better for the quantum algorithm than for
the optimal classical algorithm, 
and that the information gained by that single query is
higher.  This is true even when no entanglement is ever present
throughout the quantum computation and even when the state of the
quantum computer is arbitrarily close to being totally mixed.
The case of more than one query is left for future research,
as well as the case of a fixed \emph{average} number of queries 
rather than a fixed \emph{maximum} number. 

\section{Pseudo-Pure States}

To show that no entanglement occurs throughout our quantum computation,
we use a \mbox{special} quantum state known as
\emph{pseudo pure state} (PPS)~\cite{Chuang-Ger}.
This state occurs naturally in the framework of
Nuclear Magnetic Resonance (NMR) quantum
computing~\cite{CFH97}, but the results presented in our paper
are inherently interesting, regardless of the original NMR motivation. 
Consider any pure state $\ket{\psi}$ on $n$-qubits and some real number
\mbox{$0 \le \epsilon \le 1$}.  A~pseudo-pure state has the following form:
\begin{equation} \label{pps}
\rho_{\rm PPS}^{\{n\}}\ \equiv \ 
\epsilon |\psi\rangle\langle \psi| +
(1-\epsilon) {\cal I} \,.
\label{PPS}
\end{equation}
It is a mixture
of pure state $|\psi\rangle$ with the totally mixed state
${\cal I}=\frac{1}{2^n}\ident_{2^n}$
(where $\ident_{2^n}$ denotes the identity matrix of order $2^n$).
For example, the Werner state~(\ref{Werner}) is a special case of a PPS.

To understand why these states are called \emph{pseudo-pure}, consider
what happens if a unitary operation $U$ is performed on state
$\rho = \rho_{\rm PPS}^{\{n\}}$ from Eq.~\ref{PPS}.
\begin{proposition}\label{prop}
The purity parameter $\epsilon$ of pseudo-pure states is conserved under unitary
transformations.
\end{proposition}

\begin{proof}
Since $\rho\stackrel{U}\rightarrow U\rho U^\dag$ and
$U {\cal I} U^{\dagger}={\cal I}$,
\[ U \rho \, U^{\dagger} = 
\epsilon U |\psi\rangle\langle \psi| U^{\dagger} +
(1-\epsilon) U {\cal I} U^{\dagger} =
\epsilon |\phi\rangle\langle \phi| +
(1-\epsilon) {\cal I} \ , \]
where $\ket{\phi}=U\ket{\psi}$.
In other words, unitary operations affect only the pure part of these
states, leaving the totally mixed part unchanged and leaving the pure
proportion $\epsilon$ intact.
\end{proof}

The main interest of pseudo-pure states in our context 
comes from the fact that \emph{there exists} some bias $\epsilon$ below which
these states are never entangled.
The following theorem was originally proven in 
\mbox{\cite[Eq.~(11)]{braunstein-et-al}} but an easier proof
was subsequently given in~\cite{SchackCaves00}: 
\begin{theorem}\label{thm}
\textup{(Braunstein \emph{et al.}~\cite{braunstein-et-al})}
For any number $n$ of
qubits, a state $\rho_{\rm PPS}^{\{n\}}$ is separable whenever
\begin{equation}
\epsilon<\frac1{1+2^{2n-1}} \, ,
\label{eq:bound}
\end{equation}
regardless of its pure part~$\ket{\psi}$.
\end{theorem}

When $\ket{\psi}$ is entangled but $\rho_{\rm PPS}^{\{n\}}$
is separable, we say that the PPS exhibits \emph{pseudo-entanglement}.
(Please note that Eq.~(\ref{eq:bound}) is sufficient for separability
but not necessary.)
The key observation is provided by the Corollary below,
whose proof follows directly from Theorem~\ref{thm} and Proposition~\ref{prop}.
\begin{corollary}\label{key}
Entanglement will \emph{never} appear in
a quantum unitary computation that starts in a separable PPS whose purity
parameter~$\epsilon$ obeys Eq.~(\ref{eq:bound}).  A~final measurement in
the computational basis will not make entanglement appear either.

\end{corollary}

\section{The Deutsch-Jozsa Problem}
The problem considered by Deutsch and Jozsa~\cite{DJ92} was the following.
We are given a function $f:\{0,1\}^{n}\rightarrow \{0,1\}$
in the form of an oracle (or a black-box),
and we are promised that either this function is \emph{constant}---$f(x)=f(y)$
for all $x$ and $y$---or that it is \emph{balanced}---$f(x)=0$ on exactly half
the $n$-bit strings $x$.  Our task is to decide which is the case.
Historically, this was the first problem ever discovered for which a quantum
computer would have an exponential advantage over any classical computer,
in terms of computing time,
provided the correct answer must be given with certainty.  In~terms of the
number of oracle calls, the advantage is in fact much better than exponential:
a single oracle call (in which the input is given in superposition) suffices
for a quantum computer to determine the answer with certainty, whereas
no classical computer can be sure of the answer
before it has asked \mbox{$2^{n-1}+1$} questions.
More to the point, no information
\emph{at all} can be derived from the answer to \emph{a single} classical
oracle call. 

The quantum algorithm of Deutsch and Jozsa (DJ) solves this problem
with a single query to the oracle by starting with state $\ket{0^n}\one$
and performing a
Walsh--Hadamard transform on all $n+1$ qubits before and after the
application of $U_f$.  A~measurement of the first $n$ qubits is made
at the end (in the computational basis), yielding classical $n$-bit string~$z$.
By virtue of phase kickback, the initial Walsh--Hadamard transforms and
the application of
$U_f$ result in the following state:
\begin{eqnarray}\label{eq:dj}
|0^n\rangle |1\rangle &\stackrel{\WHs}{\longrightarrow}&
	 \bigg(\frac1{2^{n/2}}\sum_x |x\rangle \bigg)|-\rangle\ 
\stackrel{U_f}{\longrightarrow} 
\bigg(\frac1{2^{n/2}}\sum_x (-1)^{f(x)}|x\rangle\bigg)|-\rangle \ .
\end{eqnarray}
Then, if $f$ is constant, the final Walsh--Hadamard transforms 
revert the state back to
$ \pm |0^n\rangle |1\rangle $,
in which the overall phase is $+$ if $f(x)=0$ for all $x$ and
$-$ if $f(x)=1$ for all $x$.  In either case, the result of the
final measurement is necessarily $z=0$.
On~the other hand, if $f$ is balanced, the phase of half the $\ket{x}$
in Eq.~(\ref{eq:dj}) is $+$ and the phase of the other half is~$-$.
As~a result, the amplitude of $\ket{0^n}$ is zero after
the final Walsh--Hadamard transforms because each $\ket{x}$
is sent to $+\ket{0^n}/2^{n/2}+\cdots$ by those transforms.
Therefore, the final measurement can\emph{not} produce~$z=0$.
It~follows from the promise that if we obtain $z=0$ we can conclude that $f$
is constant and if we obtain $z\neq 0$ we can conclude that $f$
is balanced.
Either way, the probability of success is 1 and the
algorithm provides full information on the desired answer.

On the other hand, due to the special nature of the DJ problem, 
a single query does not change our probability
of guessing correctly whether the function is balanced or constant.
Therefore the following proposition holds:
\begin{proposition}
When restricted to a single DJ oracle call,
a classical computer learns no information about the type of $f$.
\end{proposition}
In sharp contrast, the following Theorem shows the
advantage of quantum computing even without entanglement.

\begin{theorem}
When restricted to a single DJ oracle call,
a~quantum computer whose state is never entangled can learn
a positive amount of information about the type of $f$.
\label{lemma:dj}
\end{theorem}

\begin{proof}
Starting with a PPS in which the pure part is 
$\ket{0^n}\one$ and its probability is $\epsilon$,
we can still follow the Deutsch-Jozsa strategy, but now it becomes a
guessing game. We obtain the correct answer with different probabilities
depending on whether $f$ is constant or balanced:
If $f$ is constant, we obtain $z=0$ with probability
\[P(z=0 \mid f\mbox{ is constant})=\epsilon + (1-\epsilon)/2^n\]
because we started with state $\ket{0^n}\one$ with probability
$\epsilon$, in which case the Deutsch-Jozsa algorithm is guaranteed
to produce $z=0$ since $f$ is constant, or we started with a completely
mixed state with complementary probability \mbox{$1-\epsilon$},
in which case the Deutsch-Jozsa algorithm produces a completely
random $z$ whose probability of being zero is $2^{-n}$.
Similarly,
\[P(z\neq0 \mid f\mbox{ is constant})= \frac{2^n-1}{2^n}(1-\epsilon) \, . \]
If $f$ is balanced we obtain a non-zero $z$ 
with probability 
\[P(z\ne0 \mid f\mbox{ is balanced})=\epsilon + \frac{2^n-1}{2^n}(1-\epsilon)
\, , \] and $z=0$ is obtained with probability 
\[P(z=0 \mid f\mbox{ is balanced})=(1-\epsilon)/2^n \, .  \]
For all positive $\epsilon$ and all $n$, we still observe an advantage over
classical computation. 
In~particular, this is true for
$\epsilon\leq 1/(1+2^{2n+1})$,  in which case the state remains
separable throughout the entire computation (Eq.~(\ref{eq:bound}) with $n+1$
qubits.)

Let the \emph{a priori} probability of $f$ being constant be $p$
(and therefore the probability that it is balanced is $1-p$).
The probability $p_0$ of obtaining $z=0$ is
\( \frac{1-\epsilon}{2^n}+\epsilon p. \)
We would like to quantify the amount of information we gain about the 
function, given the outcome of the measurement.
In order to do this, we calculate 
the mutual information between $X$ and $Y$, where $X$ is a random 
variable signifying whether $f$ is constant or balanced, and $Y$ is a random
variable signifying whether $z=0$ or not.
Let the \emph{entropy function} of a probability $q$ be
$ h(q) \equiv -q\log q - (1-q)\log (1-q)$.
Then, the information gained by a single quantum query is
\looseness=-1
\[
I(X;Y) = h(p)-
p_0 h\left(\frac p{p_0}
			\left(\epsilon+\frac{1-\epsilon}{2^n}\right)\right)-
(1-p_0)h\left(\frac{p(1-\epsilon)}{1-p_0}
			\left(1-\frac{1}{2^n}\right)\right) \, .
\]

\begin{figure}[tb]
\centerline{{\includegraphics[width=4in]{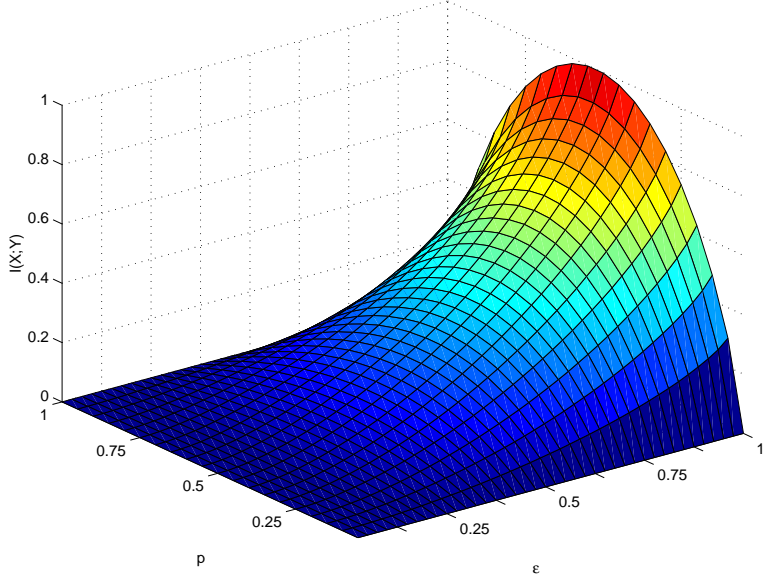}}}
\caption{\small Information gained by one quantum Deutsch--Jozsa query:
Always positive when \mbox{$0<p<1$}, even for
extremely small positive $\epsilon$}
\label{fig:inf_one_query}
\end{figure}

As shown by Figure~\ref{fig:inf_one_query}, 
the mutual information is positive for every $\epsilon > 0$, 
unless $p=0$ or $p=1$.
This is obviously more than the zero amount of information gained by a single
classical query.
For $p= 1/2$ and very small $\epsilon$ ($\epsilon\ll\frac1{2^n}$) we obtain
(cf.~Appendix~\ref{detail:dj}) that
\begin{eqnarray}\label{eq:gloup}
I(X;Y) &=&
 1-p_0h\left(\frac12+\frac{2^n\epsilon}4+O(2^n\epsilon^2)\right)
-(1-p_0)h\left( \frac12-\frac{\epsilon}{1-4/2^n}+O(2^n\epsilon^2)\right) \nonumber \\
&=&
\frac{2^{2n}\epsilon^2}{8(2^n-1)\ln2}+O(2^{2n}\epsilon^3) > 0.
\end{eqnarray}
\end{proof}

\noindent
For a specific example, consider $p=1/2$, $n=3$ and
$\epsilon=1/(1+2^{2n+1})=1/129$.  In this case, we gain
0.0000972 bits of information.

We conclude that some information is gained even for separable PPSs, 
in contrast to the classical case where the mutual 
information is always zero.
Furthermore, some information is gained even when $\epsilon$
is arbitrarily small.

We can further improve the expected amount of information that is obtained
by a single call to the oracle if we measure the $(n+1)$st qubit and take it
into account. Indeed, this qubit should be $|1\rangle$ if the
contribution comes from the pure part.  Therefore, if that extra bit
is $|0\rangle$, which happens with probability
$(1-\epsilon)/2$, we know that the PPS contributes the fully
mixed part, hence no useful information is
provided by~$z$ and we are no better than in the classical case.
However, knowing that you don't know something is better than not knowing
at all, because it makes the other case more revealing!
Indeed, when that extra bit is $|1\rangle$, 
which happens with probability $(1+\epsilon)/2$,
the probability of the pure part is enlarged from $\epsilon$ to
$\hat\epsilon = 2\epsilon/(1+\epsilon)$, 
and the probability of the mixed part
is reduced from $1-\epsilon$ to 
$1 - \hat\epsilon = (1-\epsilon)/(1+\epsilon)$.
The~probability of $z=0$ changes to
\( \hat p_0 = \frac{1-\hat{\epsilon}}{2^n}+\hat{\epsilon} p ,\)
and mutual information to
\begin{eqnarray*}
I(X;Y)&=&\frac{1+\epsilon}{2}
\left[ h(p) - \hat p_0 h \left( 
  \frac p{\hat{p_0}} \left(\hat\epsilon+\frac{1-\hat\epsilon}{2^n}\right)\right)
  -(1-\hat p_0)h\left( 
  \frac{p(1-\hat\epsilon)}{1-\hat{p_0}}\left(1-\frac{1}{2^n}\right)\right)\right]\\
\txtline{which, for $p=1/2$ and very small $\epsilon$, gives}
I(X;Y)&=&\frac{2^{2n}\epsilon^2}{4(2^n-1)\ln2}+O(2^{2n}\epsilon^3) > 0 \, .
\end{eqnarray*}
This is essentially twice as much information as in Eq.~(\ref{eq:gloup}).
For the specific example of $p=1/2$, $n=3$ and $\epsilon=1/129$, 
this is 0.000189 bits of information.

\section{The Simon Problem}
An oracle calculates a function $f(x)$ from $n$ bits to $n$ bits.
We are promised that 
$f$ is a two-to-one function, so that for any $x$ there
exists a unique $y \mbox{} \neq x$ such that $f(x) = f(y)$.
Furthermore, 
we are promised the existence of an $s \ne 0$ such that 
$f(x) = f(y)$ for $x \neq y$ if and only if
$y = x \oplus s$,
(where $\oplus$ is the bitwise exclusive-or operator). 
The goal is to find $s$, 
while minimizing the number of times $f$ is calculated.

Classically, even if one calls function $f$ exponentially
many times, say $2^{n/4}$ times, 
the probability of finding $s$ is still exponentially
small with $n$, that is less than $2^{-n/2}$. 
However, there exists a quantum algorithm that requires only $O(n)$ 
computations of $f$.
The algorithm, due to Simon~\cite{Simon97},
is initialized with $|0^n\rangle |0^n\rangle$.
It performs a Walsh--Hadamard transform on the first register and
calculates $f$ for all inputs to obtain
\begin{eqnarray}
 \ket{0^n} \ket{0^n} &\stackrel \WHs\longrightarrow &
 \frac1{2^{n/2}} \sum_x \ket x \ket{0^n} 
 \stackrel{U_f}\longrightarrow 
 \frac1{2^{n/2}} \sum_x \ket x \ket{f(x)} \, , \\
\txtline{which can be written as}
&=&\frac1{2^{n/2}} \sum_{x<x\oplus s} 
     \left(\ket x+\ket{x\oplus s}\right) \ket{f(x)}\ .\nonumber
\end{eqnarray}
Then, the Walsh--Hadamard transform is performed again on the first
register  (the one holding the superposition of all $|x\rangle$),
which produces state
\begin{eqnarray*}
\frac1{2^{n}} \sum_{x<x\oplus s}\sum_j 
   \left((-1)^{j\cdot x}+(-1)^{j\cdot x\oplus j\cdot s}\right) \ket j\ket{f(x)} \, ,
\end{eqnarray*}
where `$\cdot$' is the inner
product modulo 2 of the binary strings of $j$ and $s$.
Finally the first register is measured. Notice that the outcome $j$ is 
guaranteed to be orthogonal to $s$ ($j\cdot s=0$)
since otherwise $\ket j$'s amplitude 
$(-1)^{j\cdot x}\left(1+(-1)^{j\cdot s}\right) $ is zero.
After an expected number of such queries in $O(n)$, one obtains
$n$ linearly independent
$j$s that uniquely define $s$.

Let $S$ be the random variable that describes parameter $s$, and $J$
be a random variable that describes the outcome of a single measurement.
We would like to quantify how much information about $S$ is gained by a 
\emph{single} query.
Assuming that $S$ is distributed uniformly in the range
$[1\,.\,.\,2^n-1]$, its
entropy before the first query is \( H(S)=\log(2^n-1) \approx n\).
In the classical case, a single
evaluation of~$f$
gives no information about $S$: 
the value of $f(x)$ on any specific $x$ says nothing about its
value in different places, and therefore nothing about~$s$.
However, in the case of the quantum algorithm, we are assured that
$s$ and $j$  are orthogonal.  If the measured $j$ is zero,
$s$ could still be any one of the $2^n-1$ non-zero values and no information
is gained.  But in the overwhelmingly more probable
case that $j$ is non-zero, only $2^{n-1}-1$ values for $s$ are still possible.
Thus, given the outcome of the measurement, the entropy of $S$ drops to
approximately $n-1$ bits and the expected information gain is nearly one bit
(See Appendix~\ref{detail:simon} for a detailed calculation).
More formally, based on the conditional probability
\[P(J=j \mid S=s)=\left\{\begin{array}{ll}
	\frac{2}{2^n} & \mbox{if }j\cdot s=0\\[1ex] 
        0 & \mbox{if }j\cdot s=1 \, ,
\end{array}\right.\]
it follows that the conditional entropy $H(J \mid S=s)=n-1$, which
does not depend on the specific $s$ and therefore $H(J|S)=n-1$ as well.
In order to find the \emph{a priori} entropy of $J$, we calculate its marginal
probability
\begin{eqnarray*}
P(J=j)&=&\sum_s P(s)P(j|s) \\
&=& \left\{\begin{array}{ll}
       \frac{1-\frac{2}{2^n}}{2^n-1} & \mbox{if }j\neq0\\[1ex]
       \frac{2}{2^n} &\mbox{if }j=0\, .
    \end{array}\right.
\end{eqnarray*}
Thus,
\begin{eqnarray*}
H(J)=-\sum_jP(J=j)\log P(J=j)
&=&-\left( 1-\frac{2}{2^n}\right)\log\frac{1-\frac2{2^n}}{2^n-1}
  -\frac2{2^n}\log \frac2{2^n}\\
&=&\left( 1-\frac{2}{2^n}\right)
    \left(n+\log\frac{2^n-1}{2^n-2}\right)+\frac{n-1}{2^{n-1}}\\
\end{eqnarray*}
and the mutual information
\[
I(S;J)=1-\frac{2-(2^n-2)\log{\frac{2^n-1}{2^n-2}}}{2^n}=1-O(2^{-n})
\]
is almost one bit.

\begin{figure}[b!]
\psfrag{epsilon}{$\epsilon$}
\centerline{{\includegraphics[width=4in]{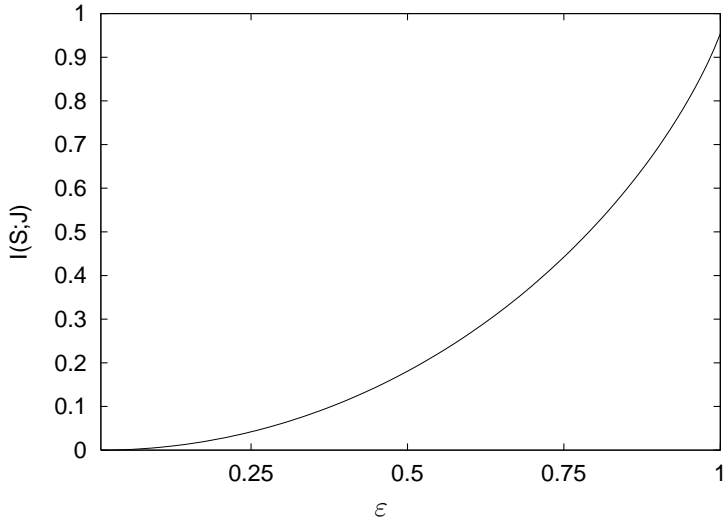}}}
\caption{Information gained by one quantum Simon query (here for $n=10$):
Always positive, even for extremely small positive $\epsilon$.}
\label{fig:inf_simon}
\end{figure}

In contrast, a single query to a classical oracle provides
no information about~$s$.
\begin{proposition}
When restricted to a single oracle call,
a~classical computer learns
no information about Simon's parameter $s$.
\end{proposition}
Again in sharp contrast, the following Theorem shows the
advantage of quantum computing without entanglement,
compared to classical computing.
\begin{theorem}
When restricted to a single oracle call,
a~quantum computer whose state is never entangled can learn
a positive amount of information about Simon's parameter $s$.
\label{lemma:simon}
\end{theorem}
\begin{proof}
Starting with a PPS in which the pure part is 
$|0^n\rangle |0^n\rangle$, and its probability is $\epsilon$,
the acquired $j$ is no longer guaranteed to be orthogonal to $s$.
In fact, an orthogonal $j$ is obtained 
with probability $\frac{1+\epsilon}2$ only.
For any value of $S$, the conditional distribution of $J$ is
\[P(J=j \mid S=s)=\left\{\begin{array}{ll}
	\frac{1+\epsilon}{2^n} & \mbox{if }j\cdot s=0\\[1ex]
        \frac{1-\epsilon}{2^n} & \mbox{if }j\cdot s=1
\end{array}\right.\]
from which we calculate (see Appendix~\ref{detail:simon}) 
that the information gained about $S$ given the value of $J$ is
\[
I(S;J)=-\left( 1-\frac{1+\epsilon}{2^n}\right)
    \log\frac{1-\frac{1+\epsilon}{2^n}}{2^n-1}
    +\left(2^{n-1}-1\right)\frac{1+\epsilon}{2^n}\log\frac{1+\epsilon}{2^n}
    +\frac{1-\epsilon}{2}\log\left( \frac{1-\epsilon}{2^n} \right).
\]
As shown in Figure~\ref{fig:inf_simon}, the amount of
information is larger than the classical zero for every $\epsilon>0$. 
\end{proof}

This theorem is true even for $\epsilon$ as small as
$1/(1+2^{2(2n)-1})$, in which case the state of the computer is never
entangled throughout the computation
by virtue of Corollary~\ref{key}.
For~example, when $n=3$ and $\epsilon=1/(1+2^{4\cdot3-1})=1/2049$,
we gain $147\times 10^{-9}$ bits of information.

\section{Conclusions and Directions for Further Research}

We have shown that quantum computing without entanglement
is more powerful than classical computing. We achieved this result by
using two well-known problems due to Deutsch-Jozsa and to Simon, and by
comparing quantum-without-entanglement to classical behaviour.
Our measure of performance was the amount of
Shannon information that can be obtained when a single
oracle query is allowed. 

In the paper~\cite{braunstein-et-al} that gave us Theorem~\ref{thm},
Braunstein, Caves, Jozsa, Linden, Popescu and Schack claimed that
``\ldots{}current NMR experiments should be considered as simulations of quantum
computation rather than true quantum computation, since no entanglement appears
in the physical states at any stage of the process''\,%
\footnote{Note, however, that later on
Schack and Caves~\cite{SchackCaves99} 
qualified their earlier claim and stated that
``\ldots{}we~\mbox{speculate} that the power of quantum-information 
processing comes not from entanglement itself, 
but rather from the information processing capabilities of
entangling unitaries.''}.
Much to the contrary, 
we showed here that pseudo-entanglement is sufficient
to beat all possible classical algorithms, which proves our point
since pseudo-entangled states are \emph{not} entangled!
In conclusion, a few final remarks are in order.
\begin{itemize}
\item
The quantum advantage that we have found is negligible (exponentially small).
A much better advantage might be obtained by increasing $\epsilon$ and
investigating the separability of the \emph{specific} states obtained
throughout the unitary evolution of the algorithms.
\item
The case of more than one query is left for a future research.
\item
The case of a fixed \emph{average} number of oracle calls,
rather than a fixed \emph{maximum} number of oracle calls, is also left
for future research. 
Indeed, it was pointed out by Jozsa that a classical strategy can easily
outperform  our unentangled quantum strategy
when solving the Deutsch-Jozsa problem if we restrict the number of
oracle calls to be 1 on the average.
For this, the classical computer tosses a coin.
With probability $\pbfrac12$, it does not query the oracle at all and
learns no information.
But otherwise, also with probability $\pbfrac12$, it queries the oracle
\emph{twice} on random inputs and learns full information---that the
function is balanced---if it obtains two distinct outputs.
This happens with overall probability $\pbfrac18$ if the \emph{a~priori}
probability of the function being balanced is~$\pbfrac12$,
which is much better than the exponentially small amount of information gleaned
from our unentangled quantum strategy after one oracle call.
\item
What is the connection between this work and quantum communication
complexity? (A~survey of this topic can be found in~\cite{survey}.)
Could quantum communication have an
advantage over classical communication even when entanglement is not used?
\end{itemize}

Let us mention two papers that appear at first to contradict our results:
Jozsa and Linden~\cite{JozsaLinden} showed that for a large class
of computational problems, entanglement \emph{is} required in order to achieve
an exponential advantage over classical computation.
Ambainis~\emph{et al.}~\cite{AmbSchulVaz00} showed that quantum
computation with  a certain mixed state, other than the pseudo-pure
state used by us, has no advantage over classical computation.
But obviously, there is no real contradiction between our paper and
these important results.  We provide a case in which there exists
a positive advantage of unentangled quantum computation over classical
computation.

\newpage

\begin{center}\LARGE\bf APPENDICES \end{center}
\appendix
\section{Details for the Deutsch-Jozsa Problem}
\label{detail:dj}
\newcommand{\constbal}[4]{
\setlength{\unitlength}{0.7cm}
\begin{picture}(10,4)(1,1)
\put(3,2){\vector(1,0){3}}
\put(3,2){\line(1,0){6}}
\put(4,1.6){#1}

\put(3,4){\vector(1,0){3}}
\put(3,4){\line(1,0){6}}
\put(4,4.2){#2}

\put(3,2){\vector(3,1){2}}
\put(3,2){\line(3,1){6}}
\put(3.5,2.5){#3}

\put(3,4){\vector(3,-1){2}}
\put(3,4){\line(3,-1){6}}
\put(7,2.5){#4}

\put(1,4){const.}
\put(1,2){bal.}
\put(9.4,4){zero}
\put(9.4,2){non-zero}
\end{picture}
}
The following two diagrams describe the probability that zero (or non-zero)
is measured, given a constant (or balanced) function, in the pure and the
totally mixed cases.

\begin{figure}[h!]
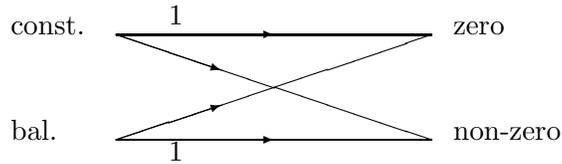

\begin{center}
\constbal{1}{1}{}{}
\end{center}
\caption{Pure initial state}
\end{figure}
\begin{figure}[h!]
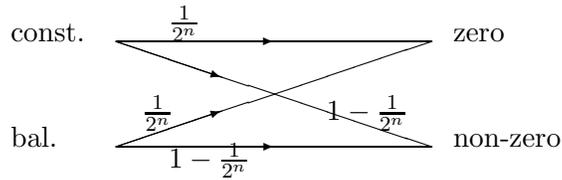

\begin{center}
\constbal{$1-\frac1{2^n}$}{$\frac1{2^n}$}{$\frac1{2^n}$}{$1-\frac1{2^n}$}
\end{center}
\caption{Totally mixed initial state}
\end{figure}

\noindent
The case of pseudo-pure initial state is the weighted sum of the previous cases.

\begin{figure}[h!]
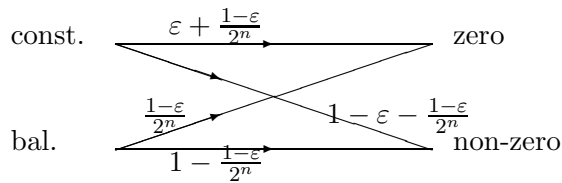

\begin{center}
\constbal{$1-\frac{1-\epsilon}{2^n}$}
         {$\epsilon+\frac{1-\epsilon}{2^n}$}
         {$\frac{1-\epsilon}{2^n}$}{$1-\epsilon-\frac{1-\epsilon}{2^n}$}
\end{center}
\caption{Pseudo-pure initial state}
\end{figure}
The details of the pseudo-pure case are summarized in the joint probability
table~\ref{tab:jointDJ} below.

\begin{table}
\begin{center}
\begin{tabular*}{3in}{p{0.55in}|p{0.9in}p{1.05in}|p{0ex}}
 \centering $X$     & \centering $y$=zero & \centering $y$=non-zero &  \\
\hline
 \centering const. & $\mystrut p\left(\epsilon+\frac{1-\epsilon}{2^n}\right)$ &
               $p(1-\epsilon)(1-\frac1{2^n})$    & \\[3ex]
 \centering bal.   & $\mxstrut (1-p)\frac{1-\epsilon}{2^n}$ & 
               $(1-p)(1-\frac{1-\epsilon}{2^n})$ & \\
\hline
 \centering $P(Y\mbox{=}y)$ &
 \mystrut $p_0=p\epsilon+\frac{1-\epsilon}{2^n}$ &
 \centering $1-p_0$ &
\end{tabular*}
\end{center}
\caption{Joint probability of function type ($X$) and measurement ($Y$)}
\label{tab:jointDJ}
\end{table}

\noindent
The marginal probability of $Y$ and $X$ may be calculated from that table,
and using Bayes rule, $P(X|Y)=\frac{P(Y|X)P(X)}{P(Y)} $, we find the conditional probabilities
\begin{eqnarray*}
P(X\mbox{=const} \mid Y\mbox{=zero})&=&\frac p{p_0}
\left(\epsilon+\frac{1-\epsilon}{2^n}\right)\\
\txtline{and}
P(X\mbox{=const} \mid Y\mbox{=nonzero})&\!\!=&
	\!\!\frac{p(1-\epsilon)}{1-p_0}\left(1-\frac{1}{2^n}\right)\\
\end{eqnarray*}
where $p_0=P(Y\mbox{=zero})=p\epsilon+\frac{1-\epsilon}{2^n}$. 
The conditional entropy is
\begin{eqnarray*}
H(X|Y) &=&\sum_y P(Y=y)h(P(X\mbox{=const} \mid Y=y))\\
&=&p_0 h\left(\frac p{p_0}
			\left(\epsilon+\frac{1-\epsilon}{2^n}\right)\right)+
(1-p_0)h\left(\frac{p(1-\epsilon)}{1-p_0}
			\left(1-\frac{1}{2^n}\right)\right),
\end{eqnarray*}
and the mutual information is, therefore,
\begin{eqnarray*}
I(X;Y)&=&H(X)-H(X|Y) \\
&=&h(p)-
p_0 h\left(\frac p{p_0}
			\left(\epsilon+\frac{1-\epsilon}{2^n}\right)\right)
-(1-p_0)h\left(\frac{p(1-\epsilon)}{1-p_0}
			\left(1-\frac{1}{2^n}\right)\right).
\end{eqnarray*}
For $p=\frac12$ this reduces into
\begin{eqnarray*}
&& 1 - \frac{1 + \epsilon\,\left( 2^{n-1}-1 \right)}{2^n} \,
     h\left(\frac{1 + \epsilon\,\left(  {2^n}-1 \right) }
       {2\,\left( 1 + \epsilon\,\left( 2^{n-1}-1 \right)  \right) }\right)\\
&&~~~- \frac{{2^n}- 1 - \epsilon\,\left( 2^{n-1}-1\right)}{2^n}\, 
h\left(\frac{\left( \epsilon-1 \right) \,\left( {2^n} -1\right) }
       {2\,\left( 1 + \epsilon\,\left( 2^{n-1}-1\right)  - {2^n} \right) }\right)
\end{eqnarray*}
and for very small $\epsilon$ ($\epsilon\ll\frac1{2^n}$) , using that fact that
$h(1/2+x)=1 - \frac{2\,x^2}{\ln2} +O(x^4)$,
this expression may be approximated by
\begin{eqnarray*}
I(X;Y)&=&1-p_0h\left(\frac12+\frac{2^n\epsilon}4+O(2^n\epsilon^2)\right)
-(1-p_0)h\left( \frac12-\frac{\epsilon}{1-4/2^n}+O(2^n\epsilon^2) \right) \\
&=&\frac{2^{2n}\epsilon^2}{8(2^n-1)\ln2}+O(2^{2n}\epsilon^3) > 0.
\end{eqnarray*}

\section{Details for the Simon Problem}
\label{detail:simon}
Let $S$ be a random variable that 
represents the sought-after parameter of
Simon's function, so that $\forall x: f(x)=f(x\oplus s)$.
Throughout this discussion, we assume that $S$ is distributed 
uniformly in the range $[1\,.\,.\,2^n-1]$.
Given that $S=s$, and starting with a PPS whose purity is $\epsilon$,
one may find the distribution of the measurement after a single
query. 
With probability $\epsilon$ 
we have started with the pure part and measured
a $j$ that is orthogonal to $s$. 
With probability $1-\epsilon$ we have started
with the totally mixed state and measured a random~$j$.
Thus for $j$ so that $j\cdot s=0$, 
\(P(J=j \mid S=s)=\epsilon\frac2{2^n}+(1-\epsilon)/{2^n},\)
and for $j$ so that $j\cdot s=1$,  
\(P(J=j \mid S=s)=(1-\epsilon)/{2^n}.\)
Putting this together,
\[P(J=j \mid S=s)=\left\{\begin{array}{ll}
	\frac{1+\epsilon}{2^n} & \mbox{if }j\cdot s=0\\[1ex]
        \frac{1-\epsilon}{2^n} & \mbox{if }j\cdot s=1 \, . 
\end{array}\right.\]
The marginal probability of $J$ for any $j\neq 0$ is
\begin{eqnarray*}
P(J=j)&=&\sum_s P(s)P(j|s) \\
&=&\frac1{2^n-1}
\left(\sum_{s\perp j}P(j|s) +\sum_{s\not\perp j}P(j|s) \right)
\\
&=&\frac{ (2^{n-1}-1)\frac{1+\epsilon}{2^n}+
2^{n-1}\frac{1-\epsilon}{2^n}}{2^n-1}\\
&=&\frac{1-\frac{1+\epsilon}{2^n}}{2^n-1},\\
\end{eqnarray*}
while for $J=0$, all values of $s$ are orthogonal, and
\begin{eqnarray*}
P(J=0)&=&\sum_s P(s)P(J=0|s) \\
&=&\frac1{2^n-1}\sum_{s\perp j}P(J=0|s)\\
&=&\frac1{2^n-1}
(2^{n}-1)\frac{1+\epsilon}{2^n}\\
&=&\frac{1+\epsilon}{2^n}.\\
\end{eqnarray*}
By definition, the entropy of the random variable $J$ is
\begin{eqnarray*}
H(J)&=&-\sum_jP(J=j)\log P(J=j)\\
&=&-\left( 1-\frac{1+\epsilon}{2^n}\right)
    \log\frac{1-\frac{1+\epsilon}{2^n}}{2^n-1}
     -\frac{1+\epsilon}{2^n}\log\frac{1+\epsilon}{2^n},\\
\end{eqnarray*}
and the conditional entropy of $J$ given $S=s$ is
\begin{eqnarray}
H(J \mid S=s) &=& 
-\sum_j P(J=j \mid S=s)\log P(J=j \mid S=s) \nonumber\\
&=&-2^{n-1}\frac{1+\epsilon}{2^n}\log\left( \frac{1+\epsilon}{2^n} \right)
-2^{n-1}\frac{1-\epsilon}{2^n}\log\left( \frac{1-\epsilon}{2^n} \right)\label{eq:Js}\\
&=&-\frac{1+\epsilon}{2}\log\left( \frac{1+\epsilon}{2^n} \right)
-\frac{1-\epsilon}{2}\log\left( \frac{1-\epsilon}{2^n} \right).\nonumber 
\end{eqnarray}
Since Eq.~(\ref{eq:Js}) is independent of the specific value $s$, it also equals to
\( H(J|S) \), which is \mbox{\( \sum_sP(S=s)H(J \mid S=s) \)}.
Finally, the amount of knowledge about $S$ that is 
gained by knowing $J$ is their mutual information:
\begin{eqnarray*}
I(S;J)&=&I(J;S)~=~H(J)-H(J|S)\\
    & = &-\left( 1-\frac{1+\epsilon}{2^n}\right)
    \log\frac{1-\frac{1+\epsilon}{2^n}}{2^n-1}
    +\left(2^{n-1}-1\right)\frac{1+\epsilon}{2^n}\log\frac{1+\epsilon}{2^n}
    +\frac{1-\epsilon}{2}\log\left( \frac{1-\epsilon}{2^n} \right).
\end{eqnarray*}
Notice the two extremes: in the pure case ($\epsilon=1$),
\(I(S;J)=1-O(2^{-n}) \)
and in the totally mixed case ($\epsilon=0$), \(I(S;J)=0\).
Finally, it can be shown that for small $\epsilon$
\[ I(S;J)=\frac{( 2^n-2 ) \epsilon^2}{2(2^n -1)\ln2}+O(\epsilon^3) \, .\]


\begin{thebibliography}{00}

\bibitem{AmbSchulVaz00}
A.~Ambainis, L.\,J.~Schulman and U.\,V.~Vazirani,
``Computing with Highly Mixed States'',
{\em \mbox{Proceedings} of the 32nd Annual ACM Symposium on the Theory of
Computing},
 697--704 (2000).

\bibitem{oracle} A.~Berthiaume and G.~Brassard,
``Oracle Quantum Computing'',
{\em Journal of Modern Optics} {\bf 41}(12), 2521--2535 (1994).

\bibitem{survey} G. Brassard,
``Quantum Communication Complexity'',
{\em Foundations of Physics}, to~appear (2003).

\bibitem{braunstein-et-al}
S.\,L.~Braunstein, C.\,M.~Caves, R.~Jozsa, N.~Linden, S.~Popescu and R.~Schack,
``Separability of Very Noisy Mixed States and Implications for NMR Quantum
Computing'',
{\em Physical Review Letters} {\bf 83}, 1054--1057 (1999).

\bibitem{BMR}
S.\,L.~Braunstein, A.~Mann and M.~Revzen,
``Maximal Violation of {B}ell Inequalities for Mixed States'',
{\em Physical Review Letters} {\bf 68}, 3259--3261 (1992).

\bibitem{CFH97}
D.\,G. Cory, A.\,F. Fahmy and T.\,F. Havel,
``Ensemble Quantum Computing by Nuclear Magnetic Resonance Spectroscopy'', 
{\em Proceedings of the US National Academy of Sciences} {\bf 94}, 1634--1639
(1997).  

\bibitem{Deutsch89} D. Deutsch,
``Quantum Computational Networks'',
{\em Proceedings of the Royal Society of London} {\bf A425}, 73--90 (1989).

\bibitem{DJ92} D. Deutsch and R. Jozsa,
``Rapid Solution of Problems by Quantum Computation'',
{\em Proceedings of the Royal Society of London} {\bf A439}, 553--558 (1992).

\bibitem{Chuang-Ger}
N.\,A. Gershenfeld and I.\,L. Chuang,
``Bulk Spin-Resonance Quantum Computation'',
{\em Science} {\bf 275}, 350--356 (1997). 

\bibitem{EkertJozsa98}
A.~Ekert and R.~Jozsa,
``Quantum Algorithms: Entanglement Enhanced Information Processing'',
{\em Philosophical Transactions of the Royal Society of London}
{\bf A356}, 1779--1782 (1998).

\bibitem{Gruska98}
J. Gruska,
\newblock {\em Quantum Computing},
\newblock McGraw-Hill, London, 1999.

\bibitem{JozsaLinden}
R.~Jozsa and N.~Linden,
``On the Role of Entanglement in Quantum Computational Speed-up'',
\texttt{arXiv:quant-ph/0201143}, 2002.

\bibitem{LindenPopescu01}
N.~Linden and S.~Popescu,
``Good Dynamics versus Bad Kinematics: Is Entanglement Needed for Quantum Computation?'',
{\em Physical Review Letters} {\bf 87}, 047901 (2001).

\bibitem{Nielsen00}
M.\,A. Nielsen and I.\,L. Chuang,
\newblock {\em Quantum Computation and Quantum Information},
\newblock Cambridge University Press, 2000.

\bibitem{SchackCaves99}
R.~Schack and C.\,M.~Caves,
``Classical Model for Bulk-Ensemble {NMR} Quantum Computation'',
{\em Physical Review~A} {\bf 60}, 4354--4362 (1999).

\bibitem{SchackCaves00}
R.~Schack and C.\,M.~Caves,
``Explicit Product Ensembles for Separable Quantum States'',
{\em Journal of Modern Optics} {\bf 47}, 387--399 (2000).

\bibitem{Shor}
P.\,W.~Shor,
``Polynomial-Time Algorithms for Prime Factorization
and Discrete Logarithms on a Quantum Computer'',
{\em SIAM Journal on Computing} {\bf 26}, 1484--1509 (1997).

\bibitem{Simon97}
D.\,R. Simon, 
``On the Power of Quantum Computation''
{\em SIAM Journal on Computing} {\bf 26}, 1474--1483 (1997).  

\bibitem{Werner_ref}
R.\,F. Werner,
``Quantum States With {E}instein-{P}odolsky-{R}osen Correlations
Admitting a Hidden-Variable Model'',
{\em Physical Review A} {\bf 40}(8), 4277--4281 (1989).


\end{thebibliography}
\end{document}